\input harvmac
\def\npb#1(#2)#3{{ Nucl. Phys. }{B#1} (#2) #3}
\def\plb#1(#2)#3{{ Phys. Lett. }{#1B} (#2) #3}
\def\pla#1(#2)#3{{ Phys. Lett. }{#1A} (#2) #3}
\def\prl#1(#2)#3{{ Phys. Rev. Lett. }{#1} (#2) #3}
\def\mpla#1(#2)#3{{ Mod. Phys. Lett. }{A#1} (#2) #3}
\def\ijmpa#1(#2)#3{{ Int. J. Mod. Phys. }{A#1} (#2) #3}
\def\cmp#1(#2)#3{{ Commun. Math. Phys. }{#1} (#2) #3}
\def\cqg#1(#2)#3{{ Class. Quantum Grav. }{#1} (#2) #3}
\def\jmp#1(#2)#3{{ J. Math. Phys. }{#1} (#2) #3}
\def\anp#1(#2)#3{{ Ann. Phys. }{#1} (#2) #3}
\def\prd#1(#2)#3{{ Phys. Rev.} {D\bf{#1}} (#2) #3}

\def\R{{\bf R}}
\def\Z{{\bf Z}}

\def\inbar{\,\vrule height1.5ex width.4pt depth0pt}
\def\IQ{\relax\,\hbox{$\inbar\kern-.3em{\rm Q}$}}
\def\IB{\relax{\rm I\kern-.18em B}}
\def\IC{\relax\hbox{$\inbar\kern-.3em{\rm C}$}}
\def\IP{\relax{\rm I\kern-.18em P}}
\def\IR{\relax{\rm I\kern-.18em R}}
\def\ZZ{\relax\ifmmode\mathchoice
{\hbox{Z\kern-.4em Z}}{\hbox{Z\kern-.4em Z}}
{\lower.9pt\hbox{Z\kern-.4em Z}}
{\lower1.2pt\hbox{Z\kern-.4em Z}}\else{Z\kern-.4em Z}\fi}

\def\n*#1{\nu^{* (#1)}}

\def\IP{{\bf P}}

\def\n#1{\nu_{#1}^*}

\def\IP{{\bf P}}

\def\n#1{\nu_{#1}^*}

\def\({ \left(  }
\def\){ \right) }

\def\-{\phantom{-}}
\noblackbox

\catcode`@=12
\baselineskip16pt
\noblackbox
\newif\ifdraft

\noblackbox
\newif\ifhypertex
\ifx\hyperdef\UnDeFiNeD
    \hypertexfalse
    \message{[HYPERTEX MODE OFF}
    \def\hyperref#1#2#3#4{#4}
    \def\hyperdef#1#2#3#4{#4}
    \def\hypernoname{}
    \def\e@tf@ur#1{}
    \def\hth/#1#2#3#4#5#6#7{{\tt hep-th/#1#2#3#4#5#6#7}}
    
\else
    \hypertextrue
    \message{[HYPERTEX MODE ON}
  \def\hth/#1#2#3#4#5#6#7{
  {\tt hep-th/#1#2#3#4#5#6#7}}

\fi

\catcode`\@=11
\newif\iffigureexists
\newif\ifepsfloaded
\def\epsfcheck{
\ifdraft
\input epsf\epsfloadedtrue
\else
  \openin 1 epsf
  \ifeof 1 \epsfloadedfalse \else \epsfloadedtrue \fi
  \closein 1
  \ifepsfloaded
    \input epsf
  \else
\immediate\write20{NO EPSF FILE --- FIGURES WILL BE IGNORED}
  \fi
\fi
\def\epsfcheck{}}
\def\checkex#1{
\ifdraft
\figureexistsfalse\immediate%
\write20{Draftmode: figure #1 not included}
\else\relax
    \ifepsfloaded \openin 1 #1
        \ifeof 1
           \figureexistsfalse
  \immediate\write20{FIGURE FILE #1 NOT FOUND}
        \else \figureexiststrue
        \fi \closein 1
    \else \figureexistsfalse
    \fi
\fi}
\def\missbox#1#2{$\vcenter{\hrule
\hbox{\vrule height#1\kern1.truein
\raise.5truein\hbox{#2} \kern1.truein \vrule} \hrule}$}
\def\lfig#1{
\let\labelflag=#1%
\def\numb@rone{#1}%
\ifx\labelflag\UnDeFiNeD%
{\xdef#1{\the\figno}%
\writedef{#1\leftbracket{\the\figno}}%
\global\advance\figno by1%
}\fi{\hyperref{}{figure}{{\numb@rone}}{Fig.{\numb@rone}}}}
\def\figinsert#1#2#3#4{
\epsfcheck\checkex{#4}%
\def\figsize{#3}%
\let\flag=#1\ifx\flag\UnDeFiNeD
{\xdef#1{\the\figno}%
\writedef{#1\leftbracket{\the\figno}}%
\global\advance\figno by1%
}\fi
\goodbreak\midinsert%
\iffigureexists
\centerline{\epsfysize\figsize\epsfbox{#4}}%
\else%
\vskip.05truein
  \ifepsfloaded
  \ifdraft
  \centerline{\missbox\figsize{Draftmode: #4 not included}}%
  \else
  \centerline{\missbox\figsize{#4 not found}}
  \fi
  \else
  \centerline{\missbox\figsize{epsf.tex not found}}
  \fi
\vskip.05truein
\fi%
{\smallskip%
\leftskip 4pc \rightskip 4pc%
\noindent\ninepoint\sl \baselineskip=11pt%
{\bf{\hyperdef\hypernoname{figure}{{#1}}{Fig.{#1}}}:~}#2%
\smallskip}\bigskip\endinsert%
}
%


\nopagenumbers\abstractfont\hsize=\hstitle
\null
\rightline{\vbox{\baselineskip12pt\hbox{EFI-97-37}
                                  \hbox{NSF-ITP-97-105}
                                  \hbox{hep-th/9708063}}}%
\vfill
\centerline{\titlefont A Note on Effective Lagrangians }
\vskip5pt
\centerline{\titlefont in Matrix Theory }
\abstractfont\vfill\pageno=0

\vskip-1.0cm
\centerline{P. Berglund$^1$ and D. Minic$^{2,3}$\footnote{$^{}$}
      {Email: berglund@itp.ucsb.edu, dminic@yukawa.uchicago.edu }}
 \vskip .20in
 \centerline{\it $^1$Institute for Theoretical Physics}          
\vskip-.4ex
 \centerline{\it University of California}         \vskip-.4ex
 \centerline{\it Santa Barbara, CA 93106, USA}       \vskip-.0ex
\vskip .05in
 \centerline{\it $^2$Enrico Fermi Institute} \vskip-.4ex
 \centerline{\it University of Chicago}         \vskip-.4ex
 \centerline{\it Chicago, IL 60637}       \vskip-.0ex
\vskip .05in
 \centerline{\it $^3$Physics Department} \vskip-.4ex
 \centerline{\it Penn State University}         \vskip-.4ex
 \centerline{\it University Park, PA 16802}       \vskip-.0ex
\vskip-.4ex
\vfill
\vskip-0.3cm
\vbox{
\baselineskip=12pt\noindent
We study the relation between the effective Lagrangian in Matrix
Theory and eleven dimensional supergravity. In particular, we provide a
relationship between supergravity operators and the corresponding terms in
the post-Newtonian approximation of Matrix Theory.}

\Date{\vbox{\line{8/97 \hfill}}}

\vfill\eject
\baselineskip=14pt plus 1 pt minus 1 pt

Matrix Theory~\ref\bfss{T. Banks, W. Fischler, S. Shenker and
 L. Susskind, {\sl M
Theory As A Matrix Model: A Conjecture},   \prd{55}
(1997) 5112, hep-th/9610043.}, the conjectured non-perturbative 
formulation
of the infinite-momentum limit of M-theory, 
even though formulated in a non-covariant,
background 
 dependent manner, seems to capture all the essential properties of 
string
 duality~\ref\banks{For a review, see T. Banks,
{\sl The State of Matrix Theory}, hep-th/9706168.}.
The comparison between Matrix Theory and its conjectured 
low-energy limit, eleven dimensional
 supergravity, in the limit of low velocities 
and large distances, has so
 far been remarkably successful \banks.
 One of the most amazing recent results is
 the fact that certain two-loop Matrix Theory effects are in complete
numerical agreement with low energy supergravity calculations
\ref\bbpt{K. Becker, M. Becker, J. Polchinski and A. Tseytlin,
{\sl Higher Order Graviton Scattering in M(atrix) Theory},
hep-th/9706072.}.
The authors of \bbpt\ developed a systematic double expansion in
relative velocity and inverse separation, the so called
post-Newtonian approximation, of the effective Lagrangian
for higher order graviton-graviton scattering in Matrix Theory.
 This approach paves
 the way for further direct comparisons of Matrix Theory predictions and
supergravity results.
It is our aim in this note to elaborate on the general structure of
effective Lagrangians in Matrix Theory in the post-Newtonian
approximation,
and further study its relationship with eleven dimensional 
supergravity
effective Lagrangian~\lref\gv{M. B. Green and P. Vanhove, 
{\sl D-instantons, Strings and
M-theory},  hep-th/9704145\semi
M. B. Green, M. Gutperle,
P. Vanhove,
{\sl One loop in eleven dimensions}, 
hep-th/9706175.}\lref\rt{J. Russo and A. A. Tseytlin, 
{\sl One-loop four-graviton
amplitude in eleven-dimensional supergravity}, 
hep-th/9707134.}~\refs{\gv,\rt}.
We work in the framework of Susskind's discrete light-cone approach to
Matrix Theory~\ref\susskind{L. Susskind, {\sl Another Conjecture about 
M(atrix)
Theory}, hep-th/9704080.}, as in~\bbpt, thus keeping the number $N$ of
D0-branes finite. (We take the light-cone coordinates to be
 defined as $x^{\pm} = t \pm x^{11}$, where $2 x^{+} = \tau$, 
the light-cone quantization
parameter, and $x^{-}$ runs from $0$ to $2\pi R$.)   
Furthermore, 
we point out that the scaling with $N$ of another remarkable two-loop 
Matrix
Theory effect, namely  scattering of gravitons off $\R^{8}/\Z_{2}$
orbifold points~\ref\ggr{O. J. Ganor, R. Gopakumar and S. Ramgoolam, 
{\sl Higher Loop
Effects in M(atrix) Orbifolds}, hep-th/9705188.}, agrees with 
eleven dimensional
supergravity.

Let us first briefly review the simple systematics of 
effective Matrix Theory Lagrangians,
following
\bbpt. Our starting point is the
 fundamental infinite-momentum frame Matrix Theory
Lagrangian ~\bfss\ (for earlier work on the $N=16$ supersymmetric
quantum mechanics, see~\ref\ch{M.~Claudson and M.~Halpern, {\sl
Supersymmetric  Ground State Wave Functions}, \npb{250} (1985) 689\semi
M. Baake, M. Reinicke and V. Rittenberg, {\sl Fierz Identities For
Real Clifford Algebras And The Number of Supercharges}, \jmp{26}
(1985) 1070\semi
R.~Flume, {\sl On Quantum Mechanics With Extended Supersymmetry And
Nonabelian Gauge Constraints}, \anp{164} (1985) 189.})
$$
S = \int dt ({ 1 \over 2R} D_{t}X_i D_t X_i + \bar{\Psi}
D_t \Psi + {1 \over 4} M^6 R [X_i, X_j]^2 - M^3 R\bar{\Psi} \gamma^i
[X_i, \Psi])\, , \eqno(1)
$$
where $X_i$ are nine $N \times N$ matrices ($N$ being kept finite), 
accompanied by sixteen supersymmetric
partners
$\Psi$, $R$ is the extent of the eleventh dimension, and $M$ is
the eleven dimensional Planck mass.
After rescaling $t R =\tau$, $M^3 X_i = y_i$, $M^3 \Psi = \psi$,
$$
S =M^{-6} \int d\tau ({ 1 \over 2} D_{\tau}y_i D_{\tau} y_i + 
 \bar{\psi}
D_{\tau} \psi + {1 \over 4} [y_i, y_j]^2 -  \bar{\psi} \gamma^i
[y_i, \psi])\, .\eqno(2)
$$
Thus, $M^6$ is the appropriate loop expansion parameter.
The effective loop action then becomes
$$
S_l= M^{6l - 6} \int d\tau f_{l} (y_i, \psi, D_\tau) = 
 M^{6l - 6}R \int dt f_{l} (M^3 X_i, M^3 \Psi, R^{-1}D_t)\, . 
\eqno(3)
$$
where $f_{l}$ is some undetermined function.
On dimensional grounds $f_l$ has to scale as $M^{8-6l}$; therefore
$$ 
S_l =  M^{6l - 6}R \int dt (M^3 r)^{4-3l}
 f_{l} ({X_i \over r},{ \Psi \over {r}}, {{D_t X_i} \over
 {RM^3 r^2}} )\, , \eqno(4)
$$
with $r^2 = X_i X_i$ and $v_{i}= D_t X_i$.
The effective loop Lagrangian at loop order $n$, $L_n$, 
written  as a function of ${v^2 \over {R^2
M^6 r^4}}$ (the powers of $v$ have to be even due to time-reversal
invariance \bbpt) then reads as follows
$$
L_n = \sum_{m=0}^\infty {v^2 \over R}
 c_{nm} ({1 \over {M^3 r^3}})^n ({v^2 \over {R^2 r^4
M^6}})^m =\sum_{m=0}^\infty {v^2 \over R}
 c_{nm} ({1 \over {r^3}})^n ({v^2 \over {r^4
}})^m ({1 \over R^2})^m ({1 \over M^3}) ^{n+2m}\, , \eqno(5)
$$
or
$$
L_n = \sum_{m=0}^\infty {v^2 \over R}
 c_{nm} ({v^2 \over {r^7}})^n ({v^2 \over {r^4
}})^{m-n} ({1 \over R^2})^m ({1 \over M^3}) ^{n+2m}\, . \eqno(5a)
$$
Here $c_{nm}$ denotes the coefficient of the $m$th term at
the $n$th loop order. Note that one can deduce the following explicit 
formula 
from the classic result of~\ref\bachas{C. Bachas, 
{\sl D-brane dynamics},
\plb{374} (1996) 37, hep-th/9511043\semi
M. R. Douglas, D. Kabat, P. Pouliot and S. H. Shenker, {\sl
D-branes and Short Distances in String Theory},
\npb{485} (1997) 85, hep-th/9608024.},   
$$
c_{1n} = { -2 \Gamma(4n+2) \over \Gamma(2n+1) \Gamma(2n+3)}
{ (-1)^{n+1} 2 (1 - 2^{2n+2})B_{n+1} - (n+1) \over 2^{4n}}\, , \eqno(6)
$$
where $B_n$ stands for the Bernoulli numbers, 
$B_n = {\Gamma(2n+1) \over \pi^{2n} 2^{2n-1}} \sum_{k=1}^{k=
\infty}{1 \over k^{2n}}$. Similarly, the tour de force calculations 
of~\ref\bb{K. Becker and
M. Becker, {\sl A Two-Loop Test of M(atrix)
Theory}, hep-th/9705091.} and \ggr\ allow
one to deduce in principle $c_{2n}$, albeit, apparently, not in a
closed form.

A very interesting physical limit of (5a) is obtained for $n=m$
$$
L_{diag} = \sum_n c_{nn} {v^2 \over R} ({v^2 \over R^2 M^9 r^7})^n\, . 
\eqno(7)
$$
This Lagrangian can be generated by considering the action of a probe 
graviton in the background of a classical source \bbpt. Note that the 
explicit powers of $N$ are not included in the above formula. In 
order to do that, one has to apply Susskind's discrete light-cone
procedure according to which the light like momentum is positive, finite 
and 
quantized as $p_{-} = N/R$~\refs{\bbpt, \susskind}. (The light like
compactification of \susskind, is to be understood as a limit of a
space-like compactification, when the space-like vector becomes null
\lref\hp{S. Hellerman and J. Polchinski, work in 
progress.}~\refs{\bbpt,\hp}.)
More explicitly, the expansion parameter in (7) comes from the action
for a massless probe in the Aichelburg-Sexl background
 in the discrete
light-cone frame (the only component of the metric induced by the
classical source  being
$h_{--} ={15 N_s \pi \over 2 R^2 M^9 r^7}$) . Then (7) can be rewritten as
$$
L_{diag} = - p_{-} {\dot{x}}^{-}\, , \eqno(8)
$$
where ${\dot{x}}^{-} = {\sqrt{1 - h_{--} v^2} - 1 \over h_{--}}$,
which follows from the classical equation of motion.
Therefore 
$c_{nn} = (-1)^{n+2}
{{1\over 2}\choose {n + 1}}(15/2)^{n}$
(for a given number of loops $n$ in the Matrix Theory effective
Lagrangian).

This result implies the following important selection rules,
$$
\Delta P(N_{s}) = 1\, ,\quad \Delta P(N_{p}) = 0\, ,\quad  \Delta n =1\, ,  
\eqno(9a)
$$
$$
\Delta P(N_{s}) = 0\, ,\quad \Delta P(N_{p}) = 0\, ,\quad  \Delta n =0\, , 
\eqno(9b)
$$
where $N_{s,p}$ denote the number of source/probe $D0$-branes, and
$P(N_{s,p})$ powers of $N_{s,p})$.
In other words, powers of $N_{s}$ come from the metric only, while
powers
of $N_{p}$ come from $p_{-}$ of the probe. (The $n$th loop effective
Lagrangian scales as $N_{s}^{n}N_{p}$, which agrees with the 
$N$ dependence in the double-line prescription for  
diagrams with $n$ loops.)

Armed with the above selection rule, we proceed to study 
 the general form of the
effective Lagrangian (5) and (5a).
First we note that the general term in (5a) can be generated by
 considering two types of
 operations, starting with the well-known ${v^4 \over r^7}$ term:
\noindent
a) horizontal ``moves'', which contribute $(m-n)$ powers of ${v^2 \over
r^4}$, and
\noindent
b) diagonal ``moves'', which give $n$ powers of ${v^2 \over
r^7}$.
The two operations can be deduced by specifying the counting of powers
of $v$ and $r$. To do so, we observe that 
in Susskind's discrete light-cone formulation of Matrix Theory~\susskind,
$$
\partial_{+} \sim v^{2}, \quad\partial_{i} \sim v,
\quad h_{--} \sim { N_s  \over  R^2 M^9 r^7}, 
\quad p_{-} \sim  {N_p}/R\, . \eqno(10)
$$
For example, ${v^4 \over r^7}$ corresponds to 
$(h_{--})_{s} (\partial_{+}\partial_{+})_p$, which in turn corresponds
to $(h_{\mu \nu})_{s} ({\cal R}_{\mu \nu})_{p}$. (Here we have separated 
operators
associated with the source, from the operators associated with the probe.)

Given the counting (10) and the selection rules (9), one can construct
the following horizontal and diagonal operations,
$$\eqalignno{{\rm Horizontal (H)}\quad
 &\rightarrow \quad (\partial^2)^{2}_{s} (\partial_{i} 
\partial_{i})_p\, ,& (11a)\cr
{\rm Diagonal  (D)}\quad
&\rightarrow \quad (h_{--})_{s} \Big({\partial_{+} \partial_{+}
\over \partial_{i} \partial_{i}}\Big)_p\, . &(11b) \cr}
$$
The form of the diagonal operator is fixed by the fact that the source
provides the classical background field and by the selection rule which 
states 
 that the power of $N_s$ can only change 
by one, after applying (11b). This in turn implies 
that
each diagonal move can have only one power of the metric and therefore, 
due to
 the fact that the indices  of the
background metric $h_{--}$ can be contracted 
only by the action of $\partial_{+}$,
associated with the probe (and suitably normalized so to get the right 
powers
of velocity), each diagonal move has to contain two powers of 
$\partial_{+}$. 
The form of the horizontal operator follows from the 
fact that the "nonlocal" part of the diagonal operator has to be canceled
in order to get an overall local operator acting on the probe. Also the
horizontal move cannot contain powers of the metric because of the 
selection
rule which states that the power of $N_{s}$ does not change, 
after an application of a horizontal move.
Thus, different operators are obtained by applying, symbolically,  $D^{n}
H^{m-n}$
on the first term in the expansion (5a).

Note that we have tacitly assumed that any
 composite operator ${\cal O}$ should be represented in the
post-Newtonian approximation as ${\cal O}_{1s} {\cal O}_{2p}$. This 
implies
that powers of velocity come only from operators acting on the probe and
powers of inverse distance come only from operators acting on the source, 
while
powers of $N_s$ are generated only by the diagonal moves.

Now we use (11) in order to construct operators that appear in the
effective eleven dimensional supergravity Lagrangian as in \rt.
For example, let us consider the operator ${\cal R}^{2i} \rightarrow
{\cal R}^i_s {\cal R}^i_p$. This
operator is generated by the $(n,m)$th term in the effective Matrix 
Theory 
Lagrangian~(5a), for $i=3(n-1) -1$ and $m = 2i-1$. In other words, the
${\cal R}^{2i}$ operator in supergravity corresponds to the $n$-th loop
${v^{2+2m} \over r^{3n + 4m}} = {v^{4i} \over r^{9i}}$ term 
in Matrix Theory. In particular for 
the 
${\cal R}^4$ term discussed in~\rt, $m=3$, $n=2$, so the ${\cal R}^4$ 
operator in supergravity
corresponds to the two-loop ${v^8 \over r^{18}}$ term in Matrix Theory.
Given the results of \bb, we see that Matrix Theory provides an
explicit prediction of the coefficient of the ${\cal R}^4$ term 
\ref\bbg{K. Becker, M. Becker and M. B. Green, work in progress.}.
Note that this term is proportional to $N_s^2$, in agreement with the
general selection rule.
Note also, that operators such as ${\cal R}^2$ and ${\cal R}^6$ do
 not correspond to any terms in the effective Matrix Theory
Lagrangian,
which is consistent with the results of~\rt.

Turning to the operator ${\cal R}^{2i+1}$ one can similarly 
show that $i=3(n-1)$ and
$m=2i+1$. Then, for example, ${\cal R}^3$ and ${\cal R}^5$ 
do not correspond to any terms in
the Matrix Theory effective Lagrangian. Thus, we find that only
operators of the form ${\cal R}^{3m-1}$, where $m=1,2,...$, are allowed, 
which
is in perfect agreement with \rt. Furthermore, from (11a),
we find that the operator creating the horizontal move has dimension
$-2\cdot 3$. We can then generalize what we mean by ${\cal R}^k$ (as was 
done
in~\rt) to also include
covariant derivatives as well as scalars made from ${\cal R}$, as long as 
the
dimension is $-2\cdot k$; the statement that only operators of the 
form ${\cal R}^{3m-1}$ are
allowed still holds.

One curious fact  that stems from the general formula for the coefficients
in the one-loop Matrix Theory Lagrangian, is that only the 
second coefficient is zero.
(The other coefficients are non-zero, which can be proven from general
properties of the Bernoulli numbers). This particular term turns out to
correspond (in accordance with our general rules)
 to $(\partial^2 {\cal R})_s (\partial_k^{2} {\cal R})_p$. 
By applying one diagonal move on
this operator and dropping the total-derivative terms, the ${\cal R}^4$ 
term
is obtained as it should.

We emphasize that the above outlined procedure cannot say anything about
numerical coefficients in front of the generated operators.

Finally, we can apply our general counting 
(in particular the counting of powers of $N$) to the problem considered 
in~\ggr,
 namely 
 the scattering of a graviton off a $\R_8/\Z_2$ orbifold 
point. We would like to 
point out that there exists no disagreement in scaling of
the effective Lagrangian for this process with $N$, between Matrix
Theory (formulated in discrete light-cone gauge ) and eleven dimensional
supergravity.

The two-loop Matrix Theory result (the gauge group is
$U(N) \times U(N)$) for scattering of a graviton
off a $\R_8/\Z_2$ fixed point is, according to \ggr, 
$$
L_{orbifold} \sim {{N^3  v^2} \over r^6}\, . \eqno(12)
$$
Note that in this case we have only one $N$, because only one
graviton scatters off a fixed point. (Hence, there are no
symmetrization factors in the Feynman diagrams as in $\R^9$~\bbpt.) 
This result agrees with the general selection rules (9) and the
expression (5a).

In the case of supergravity the authors of~\ggr\ considered the
scattering of a graviton
probe off a membrane, which couples to the
 three-form $C_3$ in eleven dimensional
supergravity. (Any process involving scattering of gravitons off each
other, e.g. scattering of a graviton off its $Z_2$ mirror,
 would necessarily go at least as $v^4$, one power of $v^2$ for
each graviton).
The classical solution representing 
a membrane in eleven dimensional supergravity is given by
the following expression, as in \ref\russo {J. Russo, {\sl BPS
Bound States, Supermembranes and T-Duality in M-Theory}, 
hep-th/9703118.} 
$$\eqalign{
ds^2 &= H^{-2/3} (-dt^2 +dy_{1}^{2} +dy_{11}^{2}) + 
H^{1/3} dx_{i}dx^{i} \cr
C_{3}&= H^{-1} dt \wedge dy_{1} \wedge dy_{11}\cr
H &= 1 + {Q \over r^{6}} 
} \eqno(13)    
$$
where $Q\sim N$, and $N$ is the number of $D0$-branes (see below).
 Then the supergravity potential can be read off from the geodesic 
equation (or from the
tree-level Feynman diagram for this process)~\ggr
$$
V(r) \sim {N^3 v^2 \over r^6}\, . \eqno(14)
$$
Here, one power of $N$ comes from the light like momentum of 
the probe while $N^2$ comes from the
membrane. We treat the  membrane as a
bound state of $N$ $D0$-branes, and therefore set the charge $Q$ 
proportional
to $N$. Another power of $N$ comes from the energy momentum tensor of 
the source which is proportional to the light like momentum of the
membrane. 
Thus the powers of $N$ agree explicitly as in the case of \bbpt.
This, in our view, 
demonstrates one more time the usefulness of the discrete
light-cone approach to Matrix Theory.

Note that the ${v^2 \over r^6}$ term appears at second loop
in the effective Lagrangian (5a). In fact, this is the very
term that can be obtained by two inverse horizontal moves from the
second term on the diagonal (the familiar ${v^6 \over r^{14}}$ term).
The overall powers of $N$ agree according to (11).

We also remark that in case the orbifold point is $\R^5/\Z_{2}$, the
one loop Matrix Theory result (the gauge group is $Sp(N)$), 
which scales as ${{N v^{2}} \over {r^3}}$ 
\lref\kr{N.~Kim and S.-J.~Rey, {\it M(atrix) Theory on $T_5/Z_2$
Orbifold and Five-Brane}, hep-th/9705132.}
\refs{\kr,\ggr}
agrees with the corresponding supergravity result (in particular
the scaling with $N$) because the energy of a
longitudinal five-brane is constant in the infinite momentum frame
\ref\brmat {T. Banks, N. Seiberg and S. Shenker, {\sl Branes
from Matrices}, hep-th/9612157.}.

Finally, one could try to repeat 
the argument presented in \bbpt\
for the metric given by (13) and generate higher order terms in
the effective Lagrangian, starting from the two-loop term ${N^3
v^2\over r^6}$. By applying the diagonal operation, which
scales as ${N^3 v^2\over r^6}$, since the source is the eleven
dimensional membrane, one would obtain terms of the form ${N^{3+l}
v^{2+2l}\over r^{6+6l}}$. However, such terms would not be allowed in the
general form of (5a). In particular, the powers of inverse distance
would not match the expansion (5a). Furthermore, by taking the
horizontal moves in the opposite direction, one can in a similar way
show that the higher loop corrections to the ${N^3 v^2\over r^6}$
are not of the form (5a). Also, each loop introduces another power of
$r^{-3}$, and is subleading. Thus, we conclude that the two-loop term
does not get further corrected (this was noticed in
\ggr). (Similar arguments can be applied to
the situation recently described in~\ref\gr{R. Gopakumar and
S. Ramgoolam,
{\sl Scattering of zero branes off elementary strings in Matrix
theory}, hep-th/9708022.}. There it is argued that
the one-loop corrected $v^2$ potential for scattering of zero branes
off elementary string in Matrix Theory is exact.)

\vskip5mm
\noindent{\bf Acknowledgments:}
The authors thank K. Becker and, in particular, J. Polchinski for useful
discussions. D.M. would also like to acknowledge the hospitality of the
ITP, Santa Barbara, where this work was initiated.
The work of P.B. was  supported in part by the National
Science Foundation grant NSF  PHY94-07194.

\vfill\eject

\listrefs
\bye